\documentclass[conference]{IEEEtran}
\IEEEoverridecommandlockouts
\usepackage{cite}
\usepackage{amsmath,amssymb,amsfonts}
\usepackage{algorithmic}
\usepackage{graphicx}
\usepackage{textcomp}
\usepackage{xcolor}
\usepackage{url}
\usepackage{url}
\usepackage{tikz}
\usepackage{booktabs}
\usepackage{multirow}
\usepackage{siunitx}
\usetikzlibrary{positioning, shapes}
\usepackage[margin=1in]{geometry}
\usepackage{listings}
\usepackage{caption}
\usepackage{courier}
\usepackage{float}
\usepackage{multicol}
\usepackage{hyperref} 
\usepackage{breakurl}
\usepackage[most]{tcolorbox}
\usepackage{graphicx}
\usepackage[subtle]{savetrees}
\lstdefinestyle{javaStyle}{
    language=Java,
    basicstyle=\ttfamily\footnotesize,
    keywordstyle=\color{blue},
    commentstyle=\color{gray},
    stringstyle=\color{teal},
    breaklines=true,
    frame=single,
    backgroundcolor=\color{white}
}

\lstdefinestyle{smaliStyle}{
    basicstyle=\ttfamily\footnotesize,
    language=Java,
    keywordstyle=\color{purple},
    commentstyle=\color{gray},
    stringstyle=\color{orange},
    breaklines=true,
    frame=single,
    backgroundcolor=\color{white}
}

\def\BibTeX{{\rm B\kern-.05em{\sc i\kern-.025em b}\kern-.08em
    T\kern-.1667em\lower.7ex\hbox{E}\kern-.125emX}}

\usepackage{xspace}
\newcommand{\toolname}{{\emph{MalLoc}}\xspace}

\usepackage{pifont} % For \ding

\newcommand{\bcircle}[1]{\ding{\numexpr181 + #1}}

\begin{document}

% \title{MalLoc: Towards Fine-grained Android Malicious Payload Localization via Large Language Models}
\title{MalLoc: Toward Fine-grained Android Malicious Payload Localization via LLMs}


% \author{\IEEEauthorblockN{Anonymous Submission}}
\author{%
  \IEEEauthorblockN{%
    \begin{tabular}{c}                                 % one centred column
      Tiezhu Sun, Marco Alecci, Aleksandr Pilgun, Yewei Song, Xunzhu Tang, \\
      Jordan Samhi, Tegawendé F.~Bissyandé, Jacques Klein\\[0.8ex]
      University of Luxembourg, Luxembourg\\
      \texttt{\{firstname.lastname\}@uni.lu}
    \end{tabular}%
  }%
}

% \author{
% \IEEEauthorblockN{
% Tiezhu Sun, Marco Alecci, Aleksandr Pilgun, Yewei Song, Xunzhu Tang, Jordan Samhi, Tegawendé F. Bissyandé, Jacques Klein
% }
% \IEEEauthorblockA{
% SnT, University of Luxembourg, Luxembourg\\
% \texttt{\{firstname.lastname\}@uni.lu}
% }
% }

% \and
% \and
% \and
% \and
% \and

\maketitle

\pagestyle{plain}

\begin{abstract}
The rapid evolution of Android malware poses significant challenges to the maintenance and security of mobile applications (apps).
Traditional detection techniques often struggle to keep pace with emerging malware variants that employ advanced tactics such as code obfuscation and dynamic behavior triggering. 
One major limitation of these approaches is their inability to localize malicious payloads at a fine-grained level, hindering precise understanding of malicious behavior. 
This gap in understanding makes the design of effective and targeted mitigation strategies difficult, leaving mobile apps vulnerable to continuously evolving threats.

To address this gap, we propose \emph{MalLoc}, a novel approach that leverages the code understanding capabilities of large language models (LLMs) to localize malicious payloads at a fine-grained level within Android malware.
% \ma{I would probably be more general here and leave the details of the evaluation for the introduction}
% To evaluate the effectiveness of \emph{MalLoc}, we developed a demonstration Android application from scratch, enabling controlled and quantitative assessment of its localization capabilities. Additionally, we apply \emph{MalLoc} to a real-world malware sample to demonstrate its practical utility in identifying and localizing malicious payloads embedded in complex, obfuscated applications.
Our experimental results demonstrate the feasibility and effectiveness of using LLMs for this task, highlighting the potential of \emph{MalLoc} to enhance precision and interpretability in malware analysis.
This work advances beyond traditional detection and classification by enabling deeper insights into behavior-level malicious logic and opens new directions for research, including dynamic modeling of localized threats and targeted countermeasure development.
\end{abstract}

\begin{IEEEkeywords}
Android Malware Analysis, Malicious Payload Localization, Large Language Models
\end{IEEEkeywords}

\section{Introduction}
Android powers billions of mobile devices worldwide~\cite{turner2025android}, enabling a vast ecosystem of applications (apps) that enhance our productivity, entertainment, and daily life. 
However, the widespread adoption and open nature of the Android platform have made it an attractive target for attackers seeking to exploit users and systems through malicious apps. 
Over time, Android malware has evolved rapidly, adopting sophisticated techniques such as code obfuscation~\cite{dong2018understanding}, dynamic behavior triggering~\cite{zhang2025dynamic}, and payload repackaging~\cite{tian2017detection} to evade traditional detection mechanisms~\cite{faruki2023survey,ruggia2024unmasking}.
These evolving threats pose critical challenges to the maintenance and evolution of secure mobile apps, requiring continuous advancements in malware analysis and mitigation strategies.

While malware detection techniques have improved in identifying whether an app is malicious or benign~\cite{arp2014drebin,wu2019malscan,daoudi2021dexray,sun2021android,sun2024detectbert,sun2024android}, and malware family classification techniques can further categorize malicious apps into known families~\cite{alswaina2020android,ding2021hybrid,kim2021efficient,freitas2022malnet,sun2025temporal}, these advances still fall short of providing actionable insights at the code level. 
Specifically, most existing models lack the capability for fine-grained localization of malicious payloads, making it difficult to accurately identify the specific code locations responsible for harmful behaviors. 
This limitation impedes the understanding of how malware operates and evolves, and it prevents researchers from extracting high-quality, behaviorally meaningful features that could strengthen detection models and improve long-term resilience.
In addition, many learning-based detection and classification models act as black boxes, offering little insight into their decision-making processes. This lack of transparency hinders analysts' ability to validate predictions, understand model limitations, and derive actionable countermeasures---ultimately limiting their utility in real-world, security-critical contexts.

% In addition, many machine learning-based malware detection and classification systems operate as opaque black boxes, providing little insight into how detection or classification decisions are made. 
% This lack of transparency limits the ability of security analysts to validate model outputs, identify blind spots, and derive actionable plans for countermeasures and model refinement. Without explainability, these models risk producing unverifiable results that undermine trust and hinder their practical deployment in security-critical environments.

A few existing works~\cite{narayanan2018multi,sun2023dexbert} have attempted to localize malicious payloads at the class level. However, they suffer from limited localization precision, making it difficult to precisely pinpoint the actual code implementing malicious behaviors. Moreover, they lack the ability to provide specific behavior descriptions that explain how the identified payloads operate. 
To the best of our knowledge, our approach \toolname is the first to explore \textbf{fine-grained} malicious payload localization along two key dimensions:
\bcircle{1} \textbf{Method-level localization}, which focuses on identifying the small executable code units (Smali methods) responsible for malicious behaviors, improving precision over class-level approaches;
% \js{the smallest unit would be an instruction right?};\tiezhu{yes, changed to "small"}
\bcircle{2} \textbf{Detailed behavioral explanations}, which provides human-readable insights into the specific actions and intent of the localized malicious payloads, supporting explainability and analyst validation.
% \dcircle{3} Evaluation on real-world malware samples, which ensures that our findings and results reflect practical  challenges and threat scenarios.
To achieve this, \toolname innovatively leverages malware family knowledge as guidance, incorporates LLM-powered semantic reasoning, and employs a two-phase approach to progressively and precisely localize malicious methods and identify their behavioral roles.

\begin{figure*}[ht]
    \centering
    \includegraphics[width=\textwidth]{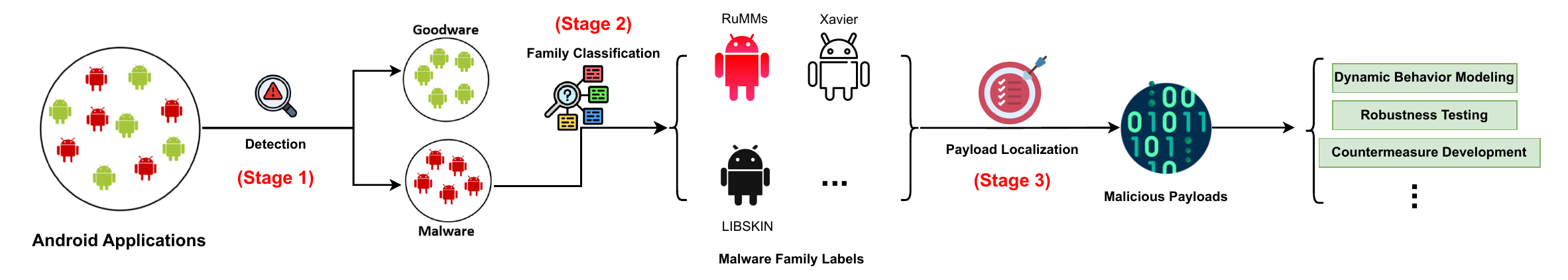}
    \caption{The overview of different stages of Android malware analysis.}
    \label{fig:analysis}
    \vspace{-1em}
\end{figure*}

% The mission of \toolname is to bridge the gap between coarse-grained malware detection or classification and the fine-grained understanding of the specific malicious behaviors embedded within malware applications.

% To enable a reliable evaluation of \toolname, we constructed a manually annotated dataset consisting of a small set of real-world Android malware samples with method-level ground-truth labels. These annotations were created with the assistance of Android security experts, who identified the specific methods responsible for malicious behaviors in each sample.
% Using this dataset, we evaluated \toolname and compared the localization performance of seven different LLMs, covering both commercial closed-source models and open-source alternatives. Our results show that XXX outperforms the others, achieving a precision of XXX and a recall of XXX in identifying malicious methods.

% \ma{Maybe add a sentence before saying that there is no ground-truth on this topic and therefore we need to rely on manual analysis --> Demo app + real-wolrd app}
% \yewei{Add a few word say it's new approach or idea} A key challenge we faced was the absence of a dataset with fine-grained ground truth for malicious behavior localization. \tiezhu{Mentioned in the last sentence of the previous paragraph}
To enable a reliable evaluation of \toolname, we developed a demo Android app, \emph{MalApp}, from scratch. It implements several common malicious behaviors observed in real-world malware, such as privacy theft and aggressive advertising, and provides fine-grained ground truth for controlled and quantitative assessment of localization capabilities.
% \js{this is contrary to what is said earlier from existing works : ``the evaluation in these studies is often based on synthesized malware samples, which do not fully reflect the complexity and diversity of real-world malware''} \js{Nevertheless you explain you have real-world apps: so I will read the rest to see how it connects.}. \tiezhu{Now I mentioned: "It implements several common malicious behaviors observed in real-world malware,"}
% \ma{maybe say that these examples are from the MalRadar paper}
In addition, we evaluate \toolname on a real-world Android malware sample from MalRadar~\cite{wang2022malradar} and manually analyze its predictions to validate their correctness.
Experimental results highlight the potential of \toolname to advance malware analysis toward more precise and explainable behavior localization. We believe this work can inform future research directions, such as dynamic analysis of localized payloads, the development of interpretable malware detection models, and the design of targeted mitigation strategies grounded in behavior-level insights.

The contributions of this work are summarized as follows:
\begin{itemize}
\item We propose \toolname, a novel LLM-driven approach for fine-grained localization of malicious payloads in Android malware, simultaneously generating corresponding behavioral explanations. %\js{and behavior description?}.
\item We develop a demo app and analyze a real-world malware sample, enabling preliminary empirical evaluation at the method level and demonstrating the potential of \toolname to advance analysis precision and explainability.
\item To support reproducibility and future research, we publicly release the dataset and source code of \toolname at: \\ ~\url{https://github.com/Trustworthy-Software/MalLoc}
\end{itemize}

\section{Background}

\noindent
\textbf{Android Malware Analysis.}
%\subsection{Android Malware Analysis}
Machine learning-based approaches~\cite{arp2014drebin,mariconti2016mamadroid,liu2024unraveling,daoudi2021dexray,10.1145/3510003.3510135,10415257} for Android malware analysis have been extensively explored over the past decade. More recently, researchers have begun investigating the potential of applying LLMs to this domain~\cite{qian2025lamd}. However, the majority of prior work remains focused on early-stage tasks---namely, malware detection (Stage 1 in Figure~\ref{fig:analysis}) and family classification (Stage 2 in Figure~\ref{fig:analysis}).
% These techniques  are crucial for identifying whether an application is malicious and assigning it to a known malware family, often leveraging static or dynamic analysis, machine learning, and heuristic-based approaches.
Despite their importance, these early-stage techniques fall short of fulfilling the ultimate goal of malware analysis: enabling effective malware defense. This includes tasks such as dynamic modeling of malicious behavior, robustness testing of detection systems, and the design of targeted countermeasures. Bridging the gap between detection/classification and actionable defense requires a deeper, more granular understanding of how malicious behaviors are implemented and triggered within an application.

A critical missing bridge in this process is fine-grained malicious payload localization (Stage 3 in Figure~\ref{fig:analysis})---the ability to identify and interpret specific methods or code segments responsible for malicious actions. Without this capability, security analysts are left with limited insight into the internal workings of malware, hindering both interpretability and mitigation efforts.
The mission of \toolname is to close this gap by advancing malware analysis beyond coarse-grained classification: \textit{ fine-grained localization and  behavioral explanation represent the core \emph{novelty} of our work}.
% By leveraging large language models (LLMs) for semantic reasoning, \toolname enables fine-grained localization and interpretation of malicious behaviors within Android applications.
\toolname paves the way for more precise, explainable, and actionable malware analysis.

\noindent
\textbf{Smali Code.}
% \subsection{Smali Code}
Android apps are primarily written in Java or Kotlin and compiled into DEX (Dalvik Executable) bytecode~\cite{dex-format}, which is stored in \texttt{.dex} files within APKs (Android Package files that bundle the compiled code and resources for distribution). % (\js{an APK file is, etc.}). 
Since these packages usually do not include the original source code, direct access to high-level code is not feasible. To enable analysis, tools such as \texttt{ApkTool}~\cite{apktool} can decompile the bytecode into Smali, a low-level, human-readable representation of DEX code. Smali serves as a practical intermediate format for examining app behavior when the original source is unavailable.
Prior work has shown that LLMs can effectively interpret and reason about Smali code~\cite{alecci2025toward}, motivating our focus on Smali-based analysis in this work.

% Alex: we don't need this example, it only eats space
% Tiezhu: This figure is for reviewers or readers who are not experts of Android. Let's decide according to the space when everything is ready.
\begin{figure*}[ht]
    \centering
    \
    \includegraphics[width=0.8\linewidth]{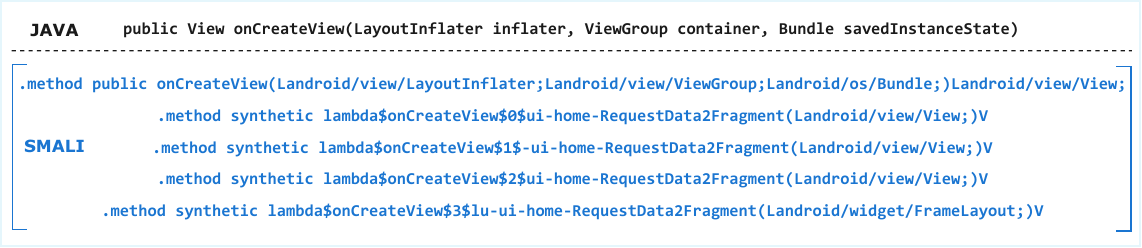}
    \caption{Example of the translation of a Java `onCreateView()` method into multiple Smali methods.}
    \label{fig:smali_method_translation}
    \vspace{-0.7em}
\end{figure*}

\begin{figure*}[ht]
\centering
\scriptsize % Apply small font to the entire figure content
\begin{minipage}{\textwidth}
\centering
\begin{tikzpicture}[
  box1/.style={
    draw, rounded corners, fill=red!5,
    text width=6.5cm, align=left, inner sep=10pt
  },
  box2/.style={
    draw, rounded corners, fill=blue!5,
    text width=3.5cm, align=left, inner sep=10pt
  },
  box3/.style={
    draw, rounded corners, fill=blue!5,
    text width=4.0cm, align=left, inner sep=10pt
  }
]

% Box 1: General Malicious Behavior Identification
\node[box1] (box1) {
\textbf{Context:} \\
You are an expert in Android malware analysis. Analyze the following Smali class and determine if it implements any malicious behaviors. \\[1ex]

\textbf{Input – Smali Class:} \\
\texttt{\{class\_content\}} \\[1ex]

\textbf{Possible Malicious Behaviors:} \\
1. Privacy Stealing; 2. SMS/CALL Abuse; 3. Remote Control; \\
4. Bank/Financial Stealing; 5. Ransom; 6. Accessibility Abuse; \\
7. Privilege Escalation; 8. Stealthy Download; 9. Aggressive Advertising; 10. Miner; 11. Tricky Behavior; 12. Premium Service Abuse. \\[1ex]

\textbf{Instruction:} \\
Use the following format: \\
\texttt{IS\_MALICIOUS: <yes or no>} \\
\texttt{CONFIDENCE: <confidence score 0-100>} \\
\texttt{EXPLANATION: <detailed explanation>} \\
\texttt{BEHAVIOR: <comma-separated behaviors>} \\[1ex]

\texttt{METHOD: <method signature>} \\
\texttt{ROLE: <role description>} \\
\texttt{METHOD: <...>} \\
\texttt{ROLE: <...>} \\[1ex]

% Do not include any other text, markdown, or formatting.
};

% Box 2: Phase 2 Behavior Detection
\node[box2, right=0.2cm of box1] (box2) {
\textbf{Context:} \\
You are an expert in Android malware analysis. Analyze the following Smali class and determine if it implements one or several of the specified malicious behaviors. \\[1ex]

\textbf{Input – Smali Class:} \\
\texttt{\{class\_content\}} \\[1ex]

\textbf{Input – Malicious Behaviors to Look For:} \\
\texttt{\{behavior\_description\}} \\[1ex] 

\textbf{Instruction:} \\
Use the following format in your response: \\
\texttt{IS\_MALICIOUS: <yes or no>} \\
\texttt{CONFIDENCE: <confidence score 0-100>} \\
\texttt{EXPLANATION: <detailed explanation>} \\[1ex]

Do not include any other text, markdown, or formatting.
};

% Box 3: Phase 1 Method Attribution
\node[box3, right=0.2cm of box2] (box3) {
\textbf{Context:} \\
The following Smali class has been identified as implementing one or several malicious behaviors in the first phase. \\
Analyze the class and identify all methods that are involved in implementing these behaviors. \\[1ex]

\textbf{Input – First Phase Explanation of Identified Malicious Behavior(s):} \\
\texttt{\{first\_phase\_explanation\}} \\[1ex]

\textbf{Input – Smali Class:} \\
\texttt{\{class\_content\}} \\[1ex]

\textbf{Instruction:} \\
IMPORTANT: For each method involved in the behavior, output the following fields, one per line, for each method: \\
\texttt{METHOD: <first line of method>} \\
\texttt{ROLE: <role description>} \\
\texttt{CONFIDENCE: <confidence score 0-100>}
};

\end{tikzpicture}
% \caption{Prompt templates used in the malicious payload localization process: Baseline Approach (left), Phase 1 of MalLoc (middle), and Phase 2 of MalLoc (right).}
\caption{Prompt templates used in: Baseline Approach (left), Phase 1 of MalLoc (middle), and Phase 2 of MalLoc (right).}
\label{fig:template}
\end{minipage}
\vspace{-1.2em}
\end{figure*}

% \ma{if it is not to difficult I would use a different color for left box (baseline) and the two right boxes (malloc)}
% \ma{also can we use subfigure here?}\tiezhu{I had trouble using subfigures. Maybe keep it like this or fix it later.}

% Smali code is the human-readable representation of Dalvik bytecode, the format used to execute Android applications.  It is generated by disassembling compiled .dex files and provides a low-level view of the instructions executed by the Android Runtime. 
During the compilation process, a single Java method can be transformed into multiple synthetic Smali methods.  This occurs when Java features like lambdas, anonymous classes, or inner classes are compiled, as they are often translated into separate methods to support efficient runtime dispatch and maintain the language's object-oriented and functional behavior within the Android environment. 
For example, in our demo app, a single Java method, \texttt{onCreateView()}, was translated into five distinct Smali methods (as shown in Figure~\ref{fig:smali_method_translation}). 
This transformation implies the challenge of mapping malicious behavior to a single Smali method.
Motivated by this observation, we design a two-phase localization approach: first identifying the malicious Smali class, then pinpointing the specific methods responsible for the behavior.
% initial analyses might benefit from a class-level perspective, as the original behavior of onCreateView() may be distributed across several Smali methods.

\section{Approach}

% \ma{Maybe here we can say we first tried something super simple, i.e., which is our baseline, and then we present MalLocl, our approach. Just to introduce the two subsections}

We begin with a simple baseline that applies LLMs directly to Smali classes. Based on its limitations (discussed later), we develop \toolname, a more structured two-phase approach. %\js{ what limitations?}
As a preprocessing step, we decompile each APK using \texttt{ApkTool}, which converts DEX bytecode into Smali format. 
%The following subsections describe the details of both approaches.
% \ma{should we mention decompilation using ApkTool as the first step common for both of them. Just to explain that we use the same prompt on each Smali Class.}

% \subsection{Baseline}
% \noindent
\textbf{Baseline.} As a starting point, we implement a straightforward baseline approach that applies LLMs directly to Smali classes, prompting the model to identify and explain malicious behaviors from a predefined list without leveraging any structural context or multi-stage reasoning.
The behavior list consists of 12 distinct malicious behaviors derived from MalRadar~\cite{wang2022malradar}, a high-quality benchmark based on real-world Android malware, with manually verified family labels. %malware family classification
In this dataset, each malware family is associated with certain particular behaviors from the list.
The behavior list and prompt template used in this baseline approach are illustrated on the left side of Figure~\ref{fig:template}.
As we demonstrate later in Section~\ref{sec:results}, the baseline approach yields poor performance. Through detailed analysis, we attribute this limitation to the inherent complexity of the task and the insufficient contextual guidance provided to the LLMs.
Specifically, the baseline prompt implicitly requires an LLM to perform three tasks simultaneously: \bcircle{1} determine whether the given Smali class is malicious or benign; \bcircle{2} identify which malicious behaviors are implemented if the class is malicious; and \bcircle{3} localize the specific methods involved in each identified behavior.
% These challenges motivate the design of our proposed method, \toolname, which we describe in the next subsection.
These challenges motivate our proposed method, \toolname, described hereafter.
% \ma{Maybe this first paragraph can be moved to the previous subsection to say something like "look this is the basic idea but it does not work too good because of these reasons --> next subsection is the solution"}

\begin{figure*}[ht]
    \centering
    \includegraphics[width=\textwidth]{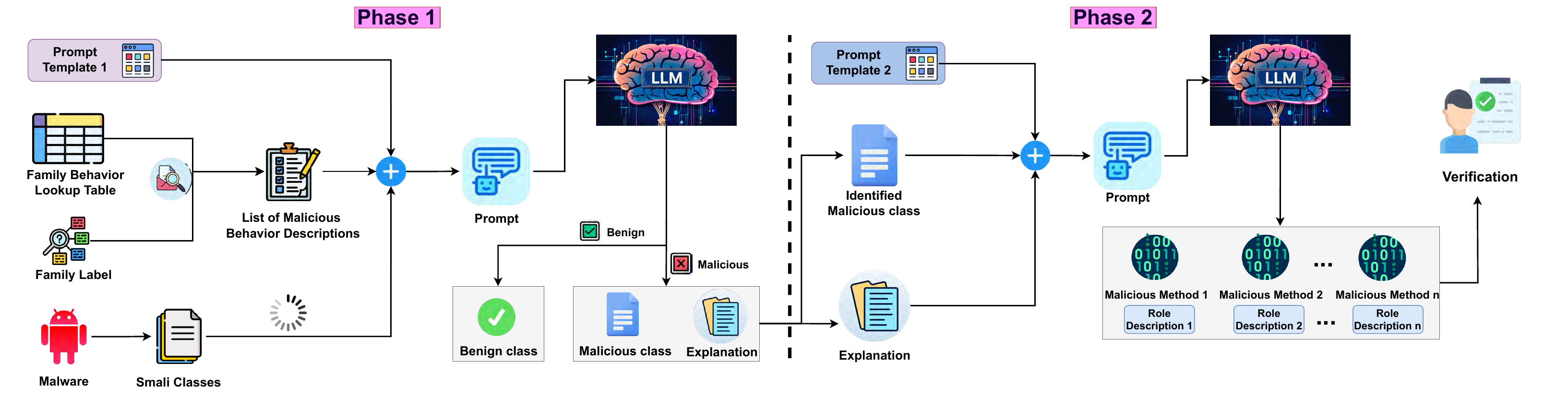}
    \caption{The overview of \toolname Pipeline. }
    \label{fig:pipeline}
    \vspace{-1em}
\end{figure*}

% \subsection{MalLoc}
% \noindent
\textbf{\toolname.} We propose \toolname, a two-phase approach that decomposes the problem along semantic and structural dimensions. Inspired by prior work showing that decomposing complex tasks can improve LLM performance~\cite{prasad2024adapt}, we isolate each subtask and enrich the prompt with targeted context. 
As illustrated in Figure~\ref{fig:pipeline}, \toolname separates localization into two distinct phases: Phase 1 focuses on identifying malicious Smali classes associated with a specific behavior, and Phase 2 drills down to pinpoint the individual methods responsible for that behavior.
This design enables \toolname to achieve more accurate and explainable localization of malicious payloads.
% \ma{do we have papers saying that decomposing the problem will bring better results? I think this can strengthen our idea of two phases approach}
% \ma{Maybe before going into details we can quickly say what each phase is doing, like o ne sentence for each phase}
% To implement \toolname, we begin by creating detailed descriptions (included in replication package) for each of the 12 predefined malicious behaviors. %\footnote{Included in:~\url{https://anonymous.4open.science/r/MalLocICSME}}
To implement \toolname, we begin by creating detailed descriptions for each of the 12 predefined malicious behaviors, which are available in the replication package.
Using metadata from MalRadar benchmark~\cite{wang2022malradar}, we construct a family behavior lookup table that maps each malware sample to a subset of relevant behaviors based on its family label. 
The two-phase pipeline is applied iteratively to each behavior linked to the input malware sample.
The prompt template for Phase 1 is shown in the center of Figure~\ref{fig:template}. It combines the Smali class content with the corresponding behavior description to prompt the LLM to determine whether the class implements that specific malicious behavior.
If a class is flagged as benign in Phase 1, the process terminates for that class. If it is deemed malicious, the class---along with the explanation generated by the LLM---is forwarded to Phase 2. As shown on the right side of Figure~\ref{fig:template}, Phase 2 uses a second prompt template that incorporates the Phase 1 explanation to guide the LLM in identifying which methods within the class contribute to the specified behavior. The output includes both the method-level localization and a role description for each method. Finally, the results are subject to manual verification by security analysts to ensure accuracy and interpretability.

\section{Experiments}
\label{sec:exps}

\subsection{Apps Under Analysis}
A key challenge in evaluating malicious behavior localization is the lack of ground-truth datasets with fine-grained annotations. 
% \ma{therefore leaving as only solution some manual evaluation, which is of course time-consuming etc... --> We need to convince the reviewer a bit why only two apps} 
In the absence of such datasets, evaluating localization accuracy requires extensive manual effort to validate behavior predictions at both class and method levels—a time-consuming and labor-intensive process. As a result, we focus our evaluation on two representative Android apps that allow for reliable and interpretable assessment. Specifically, we use: \bcircle{1} \emph{MalApp}, a controlled demo app we developed with 3 known malicious behaviors observed in real-world malware, and \bcircle{2} a real-world malware sample drawn from the MalRadar benchmark.

\subsubsection{MalApp}
This app is specifically designed to cover a diverse set of representative malicious behaviors under controlled conditions, enabling efficient verification by providing precise ground truth about where each behavior is implemented. % in the code.
The developer code of \emph{MalApp} consists of 51 Smali classes and 165 methods, among which three malicious behaviors are implemented across 3 classes and 12 methods. 
% \ma{I saw now you mention this at the end of the paragraph, but maybe it is good to anticipate here if not even in the Apps Under Analysis paragraph} 
It implements three distinct behaviors: \bcircle{1} \textit{Privacy Stealing}, \bcircle{2} \textit{Aggressive Advertising}, and \bcircle{3} \textit{Tricky Behavior}. 
For \textit{Privacy Stealing}, the app retrieves contact data from the device’s storage and transmits it to an external server. 
% \yewei{What is the input and output? Define the experiment, maybe in IV.A.} \tiezhu{the input and output are defined in the approach section. here is presenting the apps we use.}
\textit{Aggressive Advertising} is implemented via fake click generation, where a transparent overlay tricks users into interacting with an advertisement URL. %(mirroring the GhostClicker pattern).
\textit{Tricky Behavior} includes both label/icon manipulation—changing the app’s icon on user interaction—and app hiding, where the app disappears from the launcher after a trigger. 
% This modular design enables precise ground-truth labeling and straightforward manual verification.
To ensure safety and ethical compliance, all external endpoints (e.g., the server receiving contact information and advertisement links) are simulated or non-functional, and the app is intended strictly for research purposes.

\subsubsection{RuMMs App}
For our real-world analysis, we select a malware sample from the largest family in MalRadar—RuMMs—which is annotated with five malicious behaviors: \textit{Privacy Stealing}, \textit{SMS/CALL Abuse}, \textit{Remote Control}, \textit{Bank/Financial Stealing}, and \textit{Tricky Behavior}. We chose a small app in the family to minimize analysis complexity while keeping our experiments representative. The APK  contains 21 classes and 66 methods, and masquerades as an MMS/SMS messenger by using a corresponding icon and name. %is 16KB in size,
% \ma{if we put the numbers for this app we have to do it also for MalApp already in the previous section, even if you show them later}\tiezhu{}. 
The ``AndroidManifest'' lists 12 permissions—including \texttt{READ\_CONTACTS}, \texttt{SEND\_SMS}, and \texttt{BIND\_ACCESSIBILITY\_SERVICE}—and defines four activities, four services, and three broadcast receivers, suggesting a wide range of capabilities. %\texttt{CALL\_PHONE}

The app employs obfuscation techniques such as meaningless class and method names and limited reflection (e.g., classes named \texttt{a}, \texttt{b}, \texttt{Charge}, \texttt{NeglectDefend}, etc.).
Upon launch, it prompts users to enable accessibility services and attempts to become the default SMS app. It maintains persistent background services, monitors system and banking apps, and exfiltrates user contacts. Although we could not confirm the exact intent of the SMS command mechanism due to an inactive C2 server, the app clearly performs unauthorized actions like SMS sending, remote monitoring, and user tracking. 
Unlike \emph{MalApp}, reverse-engineering this malware required significant manual effort. Importantly, MalRadar provides only family-level behavior labels without payload locations or semantic explanations. Therefore, we manually verified all LLM-generated outputs to assess the accuracy and interpretability of \toolname. This real-world case allows us to test \toolname’s robustness beyond controlled environments.
Due to the time-intensive nature of such manual verification, scaling to a broader set of real-world apps is left as future work.
% \ma{maybe say here that due to the time-consuming nature, exploring other apps is left for future work}
% \ma{even if we say this later, it is just to reassure the reader}

\subsection{Results}
\label{sec:results}
\subsubsection{MalApp}

% \begin{table*}[ht]
% \centering
% \caption{Performance Comparison of Baseline and \toolname performance on \emph{MalApp}.}
% \label{tab:performance}
% \begin{tabular}{
%     ll
%     S[table-format=1.2]
%     S[table-format=1.2]
%     S[table-format=1.2]
%     S[table-format=1.2]
%     S[table-format=1.2]
%     S[table-format=1.2]
% }
% \toprule
% \textbf{Method} & \textbf{Model} & 
% \textbf{C-Precision} & \textbf{C-Recall} & \textbf{C-F1 Score} & 
% \textbf{M-Precision} & \textbf{M-Recall} & \textbf{M-F1 Score} \\
% \midrule
% \multirow{2}{*}{\textbf{Baseline}} 
%     & Phi-4    & 0.00 & 0.00 & 0.00 & 0.00 & 0.00 & 0.00 \\
%     & GPT-4.1  & 0.67 & 0.67 & 0.67 & 0.70 & 0.58 & 0.64 \\
% \midrule
% \multirow{2}{*}{\textbf{\toolname}} 
%     & Phi-4    & 0.67 & 1.00 & 0.78 & 0.62 & 0.87 & 0.70 \\
%     & GPT-4.1  & 0.83 & 1.00 & 0.89 & 0.65 & 1.00 & 0.76 \\
% \bottomrule
% \end{tabular}
% \end{table*}

%################################

\begin{table}[!hb]
\centering
\caption{Comparison of Baseline and \toolname on \emph{MalApp}.}
\label{tab:performance}
\resizebox{0.95\columnwidth}{!}{%
\begin{tabular}{
    ll
    S[table-format=1.2]
    S[table-format=1.2]
    S[table-format=1.2]
    S[table-format=1.2]
    S[table-format=1.2]
    S[table-format=1.2]
}
\toprule
\textbf{Method} & \textbf{Model} & 
\textbf{C-Prec} & \textbf{C-Rec} & \textbf{C-F1} & 
\textbf{M-Prec} & \textbf{M-Rec} & \textbf{M-F1} \\
\midrule
\multirow{2}{*}{\textbf{Baseline}} 
    & Phi-4    & 0.00 & 0.00 & 0.00 & 0.00 & 0.00 & 0.00 \\
    & GPT-4.1  & 0.67 & 0.67 & 0.67 & 0.70 & 0.58 & 0.64 \\
\midrule
\multirow{2}{*}{\textbf{\toolname}} 
    & Phi-4    & 0.67 & 1.00 & 0.78 & 0.62 & 0.87 & 0.70 \\
    & GPT-4.1  & 0.83 & 1.00 & 0.89 & 0.65 & 1.00 & 0.76 \\
\bottomrule
\end{tabular}%
} 
\vspace{-0.5em}
\end{table}

% The developer code of \emph{MalApp} consists of 51 Smali classes and 165 methods, among which three malicious behaviors are implemented across 3 classes and 12 methods. 
For our evaluation, we employed two LLMs: the open-source model Phi-4~\cite{abdin2024phi} and the commercial model GPT-4.1 from OpenAI~\cite{openai2025gpt41}.
The performance results are presented in Table~\ref{tab:performance}, where the prefix \textbf{C-} indicates class-level metrics and \textbf{M-} indicates method-level metrics.
% \ma{maybe quickly explain what we intend for TP, FP etc.. in this particular scenario. Then if we lack space we can remove it}\tiezhu{}
A prediction is a true positive (TP) if it correctly identifies a malicious class or method with the right behavior; false positives (FP) refer to incorrect behavior labels or misclassified benign components; false negatives (FN) denote missed malicious components.
% TP, FP, and FN are then used to compute the reported metrics:

% \begin{equation}
% \begin{split}
% \begin{aligned}
% \displaystyle
% \text{Precision} &= \frac{\text{TP}}{\text{TP} + \text{FP}}, \ 
% \text{Recall} = \frac{\text{TP}}{\text{TP} + \text{FN}}, \ \\ 
% \text{F1} &= \frac{2 \cdot \text{Precision} \cdot \text{Recall}}{\text{Precision} + \text{Recall}} 
% \end{aligned}
% \end{split}
% \end{equation}

The baseline approach performs poorly across both models. In particular, Phi-4 fails to produce any correct predictions. Upon manual inspection, we observed that although it successfully identified one malicious class, it misclassified the behavior type and failed to identify the relevant methods—resulting in zero scores across all metrics. While GPT-4.1 achieves slightly better performance under the baseline setup, its overall precision and recall remain low, especially at the method level.

In contrast, integrating the same LLMs into our proposed \toolname framework yields significantly improved performance. As shown in Table~\ref{tab:performance}, both models demonstrate higher precision, recall, and F1 scores at both the class and method levels. Notably, with GPT-4.1, \toolname correctly identifies all malicious classes and methods, achieving perfect recall and high precision. Specifically, it predicts 4 positive classes and 22 positive methods, compared to the total 165 methods in the app—reducing the manual analysis workload by approximately 87\%. This highlights \toolname’s strong potential to assist human analysts in accurately and efficiently localizing malicious payloads within Android applications.

\subsubsection{RuMMs App}
We apply \toolname with GPT-4.1 to the selected real-world malware sample from the RuMMs family. 
% As noted earlier, the app is disguised as an SMS messenger and uses obfuscation and accessibility abuse to conceal and enable malicious activities such as contact exfiltration, background monitoring, and SMS operations.
Due to the absence of fine-grained ground truth in MalRadar, we cannot compute all the standard class- or method-level metrics such as recall and F1 Score. However, through detailed manual analysis, we found that \toolname successfully identified 4 out of the 5 annotated behaviors associated with this sample.
Specifically, out of the app’s 21 Smali classes and 66 methods, \toolname correctly localized 6 classes and 17 methods as malicious.
All predictions and associated role descriptions were manually verified as accurate, resulting in 100\% precision at both the class and method levels.
This includes recognizing key operations such as exfiltration of contact information (\textit{Privacy Stealing}), unauthorized SMS sending (\textit{SMS/CALL Abuse}), background service persistence and remote interaction mechanisms (\textit{Remote Control}), as well as auto-clicking consent dialogs (\textit{Tricky Behavior}).
An illustrative example prediction is shown in Figure~\ref{fig:behavior-example}.

% \ma{but we can still give numbers for precision right (?)} \tiezhu{}

\begin{figure}[htbp]
\centering
\scriptsize
\begin{tcolorbox}[colback=blue!5, colframe=black!75!black,
width=0.95\columnwidth, boxrule=0.3pt, rounded corners]

\textbf{Class:} \texttt{b} (obfuscated) \\[0.5ex]
\textbf{Behavior:} \textit{Privacy Stealing}

\textbf{Method:} \texttt{.method public static f(Landroid/content/Context;)Ljava/util/ArrayList;} \\[0.5ex]
\textbf{Role Explanation:} This method enumerates the user’s contact list by querying the contacts content provider and extracting names and phone numbers.

\textbf{Method:} \texttt{.method public static a(Landroid/content/Context;ILjava/lang/String;)V} \\[0.5ex]
\textbf{Role Explanation:} This method exfiltrates the sensitive data by embedding it into Intent extras and starting a background service.

\textbf{Method:} ......
\end{tcolorbox}
\caption{An example prediction by \toolname, showing class-level behavior and method-level role explanations.}
\label{fig:behavior-example}
\vspace{-1em}
\end{figure}

These findings underscore \toolname’s ability to perform precise and semantically rich localization even in wild, obfuscated environments. 
Despite the lack of formal annotations, its outputs aligned well with expert analysis, suggesting strong potential to assist human analysts in real-world malware inspection.

% \ma{I would maybe expand on this saying indeed that, this could be anyway considered as a first step for manual analysis to reduce the search space bla bla}

% \js{intro sells two things: localization and behavior explanation. if I am not mistaken the paper only shows localization right?} \tiezhu{added an example}

% \ma{This is from RegCheck Paper, we can do something similar}

\section{Discussion}
\label{sec:discussion}

This work explores the feasibility of leveraging LLMs for the under-explored task of fine-grained malicious payload localization in Android apps. Rather than aiming for full automation, our goal is to support human analysts by narrowing the search space and reducing manual effort in identifying and understanding malicious behaviors. Our preliminary results demonstrate the promise of LLMs in this domain, particularly in enhancing interpretability and precision. At the same time, the study highlights key challenges that remain—such as improving efficiency, scaling to large apps, and calibrating confidence to guide analyst attention effectively.

\noindent
\textbf{Research Outlook.} Looking ahead, we envision that LLM-driven localization can serve as a foundation for a broad range of downstream security tasks, including dynamic behavior modeling of localized payloads, explainable malware detection, and behavior-specific mitigation strategies. We hope this work will act as a catalyst toward building more interpretable, context-aware, and analyst-assistive malware analysis systems—bridging software engineering and security, and fostering stronger synergy between LLM-based reasoning and software maintenance practices.
% \yewei{One sentence for limitations} \tiezhu{mentioned as challenges in the previous paragraph.}
\section{Conclusion}
In this paper, we presented \toolname, a novel two-phase framework that leverages LLMs for fine-grained localization of malicious payloads in Android apps. Unlike traditional work focused on coarse-grained detection or family classification, \toolname enables a deeper understanding of malicious behaviors by identifying not only the presence of malicious logic at the class level but also pinpointing the specific methods involved, along with their functional roles.

\noindent
\textbf{Future Work.} We identify several promising directions for future research. First, we plan to expand our evaluation to include additional LLMs and a more diverse set of real-world malware samples and families, in order to assess the generalizability of \toolname across behavior types and obfuscation strategies. Second, we aim to develop mechanisms for confidence calibration and uncertainty estimation to better support human analysts in high-stakes security settings. Third, we intend to explore techniques for improving the efficiency and scalability of \toolname, particularly when analyzing large and complex apps.

\section*{Acknowledgment}
This research was funded in whole or in part by the Luxembourg National Research Fund (FNR), grant references 16344458 (REPROCESS) and 18154263 (UNLOCK).

\bibliographystyle{IEEEtran}
\bibliography{reference}

\end{document}